\documentclass[11pt,a4paper]{article}

\usepackage{graphicx}
\usepackage{latexsym}

\newcommand{\bi}{\begin{itemize}}
\newcommand{\ei}{\end{itemize}}
\newcommand{\be}{\begin{equation}}
\newcommand{\ee}{\end{equation}}

\newcommand{\bea}{\begin{eqnarray}}
\newcommand{\eea}{\end{eqnarray}}
\newcommand{\beastar}{\begin{eqnarray*}}
\newcommand{\eeastar}{\end{eqnarray*}}

\newcommand{\eq}[1]{(\ref{#1})}
\newcommand{\eqq}[2]{(\ref{#1},\ref{#2})}
\newcommand{\eqqq}[3]{(\ref{#1},\ref{#2},\ref{#3})}

\title{Spontaneous relaxation in generalized oscillator models with glassy dynamics}

\author{F. Ritort\\
Departament de F\'{\i}sica Fonamental, Faculty of Physics,\\
University of Barcelona, Diagonal 647, 08028 Barcelona, Spain and\\
Department of Physics, University of California, Berkeley CA 94720, USA\\
{\tt E-Mail:ritort@ffn.ub.es}}

\begin{document}

\maketitle

\begin{abstract}
In this paper we introduce the generalized oscillator model (GOM) as a
family of exactly solvable models useful to investigate theoretical
aspects related to the statistical description of the aging
state. GOMs are defined by a potential function $V(x)$ and
characterized by a zero-temperature relaxation determined by entropy
barriers and partial equilibration. Analytic expressions for the
effective temperature can be derived using a fluctuation theorem valid
in the aging regime without the need to solve the dynamical equations
for correlations and responses. Two classes of models are investigated
in detail: the homogeneous potential model with $V(x)=(k/2p)x^{2p}$
($p$ being a positive integer) and the wedge potential model
($V(x)=k|x|$) where $V(x)$ has a singularity at the ground state
coordinate $x=0$. For the latter, we present some numerical
simulations that reinforce the validity of the main analytical
results. GOMs offer a conceptual framework to develop a statistical
description of the spontaneous relaxation process that has been
recently proposed~\cite{CriRit03b} to be at the root of the
intermittency phenomenon observed in glasses and colloids.
\end{abstract}

\section{Introduction}
Non-equilibrium phenomena is a field of research of much current
interest. From turbulence in liquids to heat convection inside stars a
plethora of systems show a very rich and complex behavior, rarely
describable in terms of few variables~\cite{CasJou03}. The opposite
situation is encountered in equilibrium systems. There few parameters
are needed to characterize the equilibrium state and its fluctuations.
Entropy, a key concept in thermodynamics, admits a statistical
interpretation in terms of the microscopic motion of
molecules. Boltzmann established the bridge linking the microscopic
and the macroscopic worlds, the central result in his theory being the
relation $S=k_B\log(W)$ where $W$ is the number of configurations
available to the system. The extension of this approach to
non-equilibrium systems and the characterization of their behavior in
terms of a few number of parameters still represents a major
theoretical challenge.

Two categories of non-equilibrium systems have received considerable
attention in past years: systems in steady states and glassy systems.
The first category encompasses all those systems driven out of
equilibrium to a stationary state by the action of an external
perturbation. The most common example is a wire of metal with extremes
in contact with two thermal sources at different temperatures. In this
case, and if the temperature difference is not too large, the flow of
heat from the hotter to the colder source is described by the Fourier
law. The second category encompasses all systems that are not in a
stationary state but which properties change very slowly with
time. Structural glasses (such as ordinary window glass) are prototype
examples.

The glass state is characterized by a very slow relaxation towards
equilibrium and by a exceedingly low rate of the energy released from
the system to the bath during the relaxation. A useful parameter to
characterize the glass state is the age of the glass (also called
waiting time in several experimental protocols) which is the time
elapsed since the system was prepared in the non-equilibrium state. In
the aging state correlation functions tend to decay in a timescale
that is of the order of the age of the system. A statistical
interpretation of the aging state has not yet been accomplished,
however our present understanding might not be far from resolving
several of the most important clues. Recent experiments have observed
the existence of intermittent fluctuations~\cite{Cil03,Cip03} that
could be related to dissipative processes characteristic of the glass
state~\cite{CriRit03b,Ritort03b}. A thermodynamic description of such
processes could provide an important step in that direction.

Statistical models have been always an important source of inspiration
and ideas to understand the glass transition.  Many types of models have
been considered in the past, from phenomenological two-state systems to
spin glasses, passing through a wide range of family systems such as
hard-spheres liquids, Lennard-Jones systems, lattice models, kinetically
constrained models, among others. Most of these models have been
investigated using approximate methods or numerical simulations.

To this list we should add exactly solvable models. From a historic
point of view, these have played an important role in the early days of
statistical mechanics. From the urn models introduced by the
Ehrenfests aiming to understand concepts such as entropy and
thermal equilibrium, to the Ising model that describes phase
transitions and critical phenomena, solvable models offer conceptual
frameworks to contrast ideas and check their consistency by evaluating
specific predictions. In this way, exactly solvable models have also
contributed to our current understanding of glassy systems.

The goal of this paper is to introduce a general family of exactly
solvable models that might help to better understand the mechanisms
behind the slow relaxation observed in glassy systems. We introduce the
generalized oscillator model (GOM) as a generalized version of a
previous model introduced by the author~\cite{BonPadRit97}. These share
some properties with kinetically constrained models as statics is
trivial but dynamics is not. Therefore they belong to a large category
of models whose dynamics can be very rich despite of the fact that the
energy landscape has trivial properties. The main objective all along
this paper is to discuss a statistical approach to the aging state by
emphasizing the relevance of the concept of effective temperature~\cite{CugKurPel97} 
as a useful way to quantify violations of the
fluctuation-dissipation theorem~\cite{CugKur93,CriRit03}.  In the framework of
the GOM we stress the relevance of the effective temperature to quantify
the spectrum of intermittent energy fluctuations in the aging regime
that have been experimentally observed~\cite{Cil03,Cip03}. The link between intermittent
effects in non-equilibrium systems and statistical effective
temperatures has been recently proposed in the framework of simple
models for the glass transition~\cite{CriRit03b,Ritort03b}. The present
paper extends these considerations to the GOM.


\section{The generalized oscillator model (GOM)}
\label{sec:gom}
Generalized oscillator models (GOMs) consist of a set of non-interacting
oscillators each described by a continuous variable $x_i$ and the energy function,
\be
E=\sum_{i=1}^N V(x_i)\qquad,
\label{gom1}
\ee
where $V(x)$ is a real valued potential energy function that diverges
to $+\infty$ in the limit $|x|\to\infty$. For instance, the potential
can be of the type $V(x)=\frac{k}{2p}x^{2p}$ with $p$ an integer
value. This is called the homogeneous potential model, the case $p=1$
corresponding to the harmonic case introduced
in~\cite{BonPadRit97}. Here we will only deal with potential energy
functions such that the partition of an individual oscillator, ${\cal
Z}_1$, remains finite at finite temperatures,
\be
{\cal Z}_1=\int_{-\infty}^{\infty} \exp\Bigl( -\beta V(x)\Bigr)\qquad,
\label{gom2}
\ee
with $\beta=1/k_BT$, $k_B$ being the Boltzmann constant (that we will
set equal to one) and $T$ is the temperature of the bath with which the
system is put in contact.

We consider a dynamics where all oscillators are updated
in parallel according to the rule 
\be
x_i\to x_i+\frac{r_i}{\sqrt{N}}\qquad,
\label{gom3}
\ee
and the $r_i$ are uncorrelated Gaussian variables with $\overline{r_i}=0$
and variance $\overline{r_i r_j}=\Delta^2\delta_{ij}$. The updating of
all oscillators is carried out in parallel in a single move. The 
move is accepted according to the Metropolis rule. We will focus our analysis on the
zero-temperature dynamics as this is the case where relaxation is fully
determined by entropic effects. Indeed, at $T=0$ activated jumps over
energy barriers are suppressed and relaxation proceeds only through search of
favorable directions in phase space where the energy decreases. As time
goes on, dynamics slows down as most of the proposed moves tend to push
the system away from the ground state by increasing its energy. Only those
moves that decrease the energy are accepted. Glassy dynamics in the GOM is then
consequence of the quick decrease with time of the acceptance rate.

The GOM shares with kinetically constrained
models~\cite{RitSol03} the property that, while statics is trivial, dynamics is
complicated. Oscillators in \eq{gom1} are non-interacting and therefore the model
has trivial static properties, the $N$-oscillator partition function
being given by ${\cal Z}_N=({\cal Z}_1)^N$. The thermodynamic properties
are then derived by computing the integral \eq{gom2}.

To solve the dynamics of the GOM we have to compute the probability
distribution of energy changes $P(\Delta E)$. This is defined as the
probability density that in a given move the energy changes by an amount
$\Delta E$. In general, this quantity shows a
complicated dependence on the $N$-oscillator probability density ${\cal
P}_t(\lbrace x_i\rbrace)$ that the system occupies the configuration
$\lbrace x_i\rbrace$ at time $t$. However, the GOM has the good property
that the $P(\Delta E)$ depends on a finite number of observables rather
than on the full configurational probability distribution. This
property is characteristic of mean-field systems, the GOM being just
another example. This makes
the model amenable of analytical computations and a good {\em laboratory} to
test many results regarding the glassy regime. To compute the probability
$P(\Delta E)$ we consider the change of energy in an elementary move,
\be
\Delta E=\sum_{i=1}^N\Bigl[V\bigl(x_i+\frac{r_i}{\sqrt{N}}\bigr)-V(x_i)\Bigr]=
\sum_{k=1}^{\infty}\frac{1}{k! N^{k/2}}\sum_{i=1}^NV^{(k)}(x_i)r_i^k\qquad.
\label{gom4}
\ee
As dynamics is stochastic the quantity $\Delta E$ is a random variable
 whose distribution can be reconstructed from the moments
 $\overline{(\Delta E)^k}$.  An explicit calculation of such moments
 shows that only the first two moments give a finite contribution in the
 large $N$ limit. Therefore, in the thermodynamic limit, $P(\Delta E)$
 is a Gaussian distribution
\be P(\Delta E)=\bigl(2\pi \sigma_{\Delta E}^2\bigr)^{-\frac{1}{2}}
      \exp\left[-\frac{(\Delta E-M_{\Delta E})^2}{2\sigma_{\Delta E}^2}\right]\qquad,
\label{gom5}
\ee
with mean $M_{\Delta E}$ and variance $\sigma_{\Delta E}^2$ given by,
\bea M_{\Delta E}=\overline{\Delta
E}=\frac{\Delta^2}{2}\overline{V^{''}(x)}\label{gom6a}\qquad ,\\
\sigma_{\Delta E}^2=
\overline{(\Delta E)^2}-(\overline{\Delta
E})^2=\Delta^2\overline{(V^{'}(x))^2}\qquad,
\label{gom6b}
\eea
where $\overline{f(x)}=(1/N)\sum_{i=1}^Nf(x_i)$. At $T=0$ the equations
for the acceptance $a$ (i.e. the fraction of accepted moves) and the energy
per oscillator $e=E/N=\overline{V(x)}$ can be written as,
\bea
a=\int_{-\infty}^{0}P(y)dy\qquad,\label{gom7a}\\
\frac{\partial e}{\partial t}=\int_{-\infty}^{0}yP(y)dy\qquad .\label{gom7b}
\eea
Inserting \eq{gom5} in \eqq{gom7a}{gom7b} we obtain,
\bea
\frac{\partial e}{\partial t}=-\Bigl(\frac{\sigma_{\Delta
E}^2}{2\pi}\Bigr)^{\frac{1}{2}}\exp\Bigl(-\frac{M_{\Delta E}^2}{2\sigma_{\Delta E}^2}\Bigr)
+\frac{M_{\Delta E}}{2}Erfc\Bigl(\frac{M_{\Delta E}}{(2\sigma_{\Delta E}^2)^{\frac{1}{2}}} \Bigr)\qquad,
\label{gom8a}\\
a=\frac{1}{2}Erfc\Bigl( \frac{M_{\Delta E}}{(2\sigma_{\Delta
E}^2)^{\frac{1}{2}}} \Bigr)\qquad,\label{gom8b}
\eea
with $Erfc(x)=(2/\sqrt{\pi})\int_x^{\infty}\exp(-u^2)du$ the
complementary error function.  These equations are not generally
solvable as they are not closed and the time evolution for the mean
$M_{\Delta E}$ and the variance $\sigma_{\Delta E}^2$ is unknown. Only
for some types of potentials $V(x)$ a closed solution can be found.
Oscillator models have the interesting property that relaxation at
$T=0$ gets steadily slower as the ground state is approached. In fact,
the ground state configuration $\lbrace x_i^{\rm GS}\rbrace$ is
characterized by the fact that the energy \eq{gom1} is an absolute
minimum, therefore $x_i^{\rm GS}=x^{\rm GS}$, with $V'(x^{\rm GS})=0$.
According to \eq{gom6b} the value of $\sigma_{\Delta E}^2$ steadily
decreases as the ground state is approached. From \eqq{gom8a}{gom8b}
this leads to a quick decrease of the decay rate of the energy and the
acceptance rate.  In such conditions the scenario of partial
equilibration (as described in \cite{Ritort03b} and below in Sec.~\ref{partial}) holds.  A salient
feature of \eqq{gom8a}{gom8b} is the dependence of the dynamical
equations upon the following parameter,
\be
\lambda=\frac{\sigma_{\Delta E}^2}{2 M_{\Delta E}} \qquad,
\label{gom9}
\ee
which has the dimensions of an energy (or temperature). Indeed, at
$T=0$ in the large-time limit, the quantity $\lambda$ vanishes
asymptotically. If we define $x=\sqrt{M_{\Delta E}/(4\lambda)}$ we can
then expand the complementary error function in \eqq{gom8a}{gom8b}
around $x=\infty$,
\be
Erfc(x)=\frac{\exp(-x^2)}{\sqrt{\pi}x}\Bigl(1-\frac{1}{2x^2}+{\cal
O}\bigl( \frac{1}{x^4}\bigr)\Bigr)~~~.
\label{gom9b}
\ee
Substituting this expansion in \eq{gom8a} we get for the time evolution
of the energy,
\be
\frac{\partial e}{\partial t}=-\Bigl(\frac{\sigma_{\Delta
E}^2}{8\pi}\Bigr)^{\frac{1}{2}}\frac{\exp(-x^2)}{x^2}\qquad.
\label{gom9c}
\ee
This equation can be asymptotically solved for a quite broad family of
models. In general, the asymptotic decay of the energy can be expressed
in terms of the parameter $\lambda$ \eq{gom9} by knowing the analytic
behavior of $V(x)$ in the vicinity of $x=0$.  We show below
how the parameter $\lambda$ in \eq{gom9} plays the role of an effective
temperature that quantifies violations of the fluctuation-dissipation
theorem~\cite{CriRit03}.

\section{Partial equilibration}
\label{partial}
\begin{figure}
  \centering \includegraphics[scale=0.3]{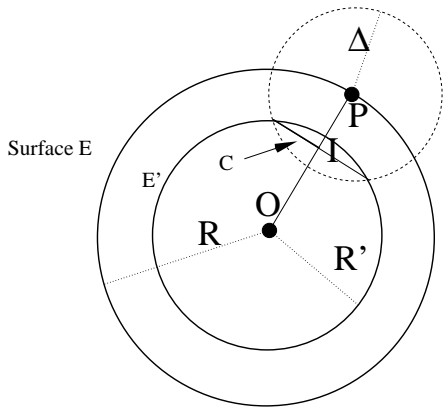} 
  \caption{Geometrical
  construction to compute $P(\Delta E)$. The thick lines denote the
  departing and final energy hypersurfaces centered at $O$. The dashed
  circle indicates the hypersurface accessible from point $P$. The intersecting
  region between the accessible hypersphere centered at $P$ and the final
  hypersurface of energy $E'$ defines a hypersurface $I$ of radius $C$
  (the radius is represented by a thick line). See
  the text for a more detailed explanation.}  
  \label{fig1}
\end{figure}

To better understand what partial equilibration means we will consider
the case of an harmonic well $V(x)=(1/2)kx^2$ where $k$ is the stiffness constant
of the well. This case corresponds to the linear harmonic oscillator
introduced in~\cite{BonPadRit97} and studied in detail in other
works~\cite{Nieuwenhuizen98,CriRit03c}.  For the harmonic case the energy is
quadratic in the variables $x_i$,
\be
E=\frac{k}{2}\sum_{i=1}^Nx_i^2
\label{gom10}
\ee
Eqs.~\eqq{gom6a}{gom6b} give $M_{\Delta E}=k\Delta^2/2,\sigma_{\Delta
E}^2=k^2\Delta^2\overline{x^2}=2k\Delta^2 E/N=2k\Delta^2 e$ where
$e=E/N$ is the energy per oscillator.  We now follow the discussion presented
in Ref.~\cite{CriRit03c}. The constant energy surface can be represented
by an hypersurface centered around the origin $x_i=x^{\rm GS}=0$
(depicted as $O$) of radius $R=\sqrt{2E/k}$. In Figure \ref{fig1} we
depict a schematic representation of the motion of a representative
configuration $\lbrace x_i^0\rbrace$ (depicted as $P$) of energy $E$ in
phase space. The smaller dashed circle represents the region of points
accessible from $\lbrace x_i^0\rbrace$ according to the dynamics
\eq{gom3}. All accessible points $\lbrace x_i \rbrace$ satisfy
$\sum_i(x_i-x_i^0)^2=\Delta^2$, i.e. lie at a distance $\Delta$ from
$\lbrace x_i^0\rbrace$ which is the radius of the smaller dashed circle.
The accessible configurations in a single move lie in a spherical
hypersurface of dimension $N-2$ corresponding to the intersection of the
hypersurface of energy $E'$ and the smaller spherical hypersurface of
radius $\Delta$. We call this region the intersecting region $I$ as
shown in Figure \ref{fig1}. The final configurations contained in $I$
lie at a distance $R'=\sqrt{2E'/K}$ to the origin $O$. The change in
energy associated to this transition is $\Delta E=E'-E$. The probability
of this jump is therefore proportional to the surface of the
intersecting region, $P(\Delta E)\propto C^{N-2}$ where $C$ is the
radius of the intersecting region.  The computation of $C$ is quite
straightforward as can be deduced from the triangle including the points
$P,O$ as vertexes and whose three sides are $R,R',\Delta$. In terms of
$R,R'$ and $\Delta$, the distance $C$ is given by the relation: 
$R=\sqrt{R'^2-C^2}+\sqrt{\Delta^2-C^2}$. Expressed in terms of
$E,\Delta E,\Delta$ we have,
\begin{equation}
  C^2=\Delta^2-\frac{k}{8E}\left( \frac{2\Delta E}{k}-\Delta^2
                           \right)^2\qquad .
\label{gom11}
\end{equation}
The surface  $\Omega(E,\Delta E)$ corresponding to the
region $I$ of radius $C$, relative to the energy $E$ of the reference
configuration $\lbrace x_i^0\rbrace$, is
\begin{equation} 
  \Omega(E,\Delta E)\propto C^{N-2} = 
              \left[\Delta^2-\frac{k}{8E}\left(
 		\frac{2\Delta E}{k}-\Delta^2\right)^2
	      \right]
	      ^{\frac{N-2}{2}}\qquad .
\label{gom12}
\end{equation}
Using the fact that $E$ is extensive with $N$ this expression can be
rewritten as,
\begin{equation}
  \Omega(E,\Delta E)\propto \exp\left[ 
             -\frac{(\Delta E-\frac{K\Delta^2}{2})^2}{4(E/N)K\Delta^2} 
	     \right]\qquad ,
\label{gom13}
\end{equation}
which is then proportional to the probability distribution \eq{gom5}. 

\begin{figure}
  \centering \includegraphics[scale=0.3]{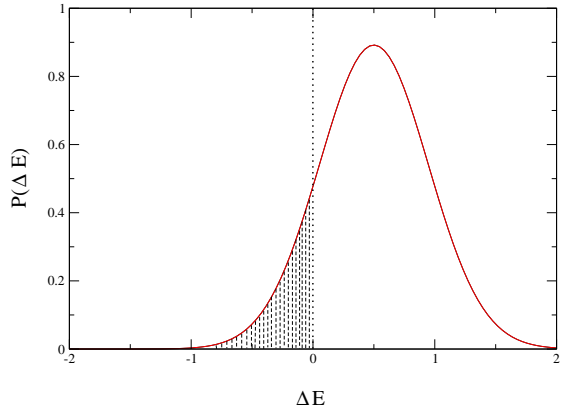} 
  \caption{Probability distribution of energy changes for the GOM
  as derived from \protect\eq{gom5} and \protect\eq{gom13}. The
  major part of energy changes are with $\Delta E>0$. At $T=0$ the variance of
  distribution $\sigma_{\Delta E}^2$ decreases relative to the mean
  $M_{\Delta E}$ (i.e. the parameter $\lambda$ in \eq{gom9} decreases as
  the ground state is approached). The fraction of accepted moves (dashed
  area with $\Delta E<0$) steadily decreases with time.}  
  \label{fig2}
\end{figure}

This construction then provides a geometric way to determine the
Gaussian distribution \eq{gom5}. From \eq{gom11} we see that $C$ (and
therefore also $\Omega(E,\Delta E)$ or $P(\Delta E)$) has a maximum
for $\Delta E=k\Delta^2/2$. The Gaussian distribution is depicted in
Fig.~\ref{fig2}. Consider now a $T=0$ dynamics where only moves with
$\Delta E<0$ are accepted. In this case, as the relative number of
configurations that are accessible from $P$ goes like $C^{N-2}$, the
largest number of accessible configurations lie in the vicinity of
$P$. Because the radius of the small dashed hypersphere in
Fig.~\ref{fig1} is equal to $\Delta$ but the radius $R$ of the
hypersurface of energy $E$ is proportional to $N^{1/2}$, dynamics constraints the
system to move along the constant energy hypersurface in the
thermodynamic limit . The system has then time to diffuse throughout a
given energy shell of finite width before leaving that shell towards
lower energy surfaces. This scenario was called partial
equilibration in \cite{Ritort03b}.

This derivation can be generalized to the GOM \eq{gom1} where the
constant energy hypersurface is not necessarily a sphere. The crucial
point in the argument is then the fact that $\Delta$ is a finite quantity
while the typical spatial dimensions of the constant energy
hypersurface are of the order $\sqrt{N}$. In the thermodynamic limit
the hypersurface is locally a sphere and the length $C$ \eq{gom11} can
be mathematically expressed in terms of the mean energy curvature
$\overline{V''(x)}$ and the modulus of the energy gradient vector
given by $\overline{V'(x)^2}$. The final result is again the most
general expressions \eqqq{gom5}{gom6a}{gom6b}.

\section{The fluctuation theorem (FT) and effective temperatures}
In the GOM a scenario of partial equilibration takes place in the
constant energy hypersurface. This is exemplified in Fig.~\ref{fig1}
for the harmonic case where it is shown how the system is constrained
to dwell in the constant energy hypersurface before moving to lower
energy configurations. In this case, because
dynamics is microscopically reversible and ergodic along the constant
energy hypersurface, a quasi-stationary dynamics emerges with a
probabilistic description that can be done in terms of a
microcanonical measure. However, there are important differences
between this dynamical measure and the usual microcanonical measure
for equilibrium systems. While in the latter the energy is kept
strictly constant and there is no time dependence of the microstate
probability distribution, in the former the measure is dynamical and
time dependent.

A key concept in such description is the notion of configurational
entropy (sometimes also called complexity) $S(E)$ which counts the
number of configurations with a given energy $E$. This quantity is
precisely given by $P(\Delta E)$ except for the fact that $P(\Delta E)$
counts the number of configurations of energy $E'$ which are accessible
from a reference configuration of energy $E$.
In a microcanonical description
of the aging state the probability to visit
configurations with energy $E'$
is then proportional to the number of configurations with that energy $\Omega(E')$,
\be
P(\Delta E)\propto \Omega (E')=\exp\Bigl( S(E') \Bigr)\qquad ,
\label{gom14}
\ee
with $\Delta E=E'-E$, $E'$ being the final energy and $E$ the departure
energy. It is important to emphasize that $P(\Delta E)$ in
\eq{gom14} (as well as in \eq{gom5}) is a probability rather than
a rate. Transition rates are transition probabilities per unit of time,
therefore have natural dimensions of frequency. We will denote them by
$W(\Delta E)$. In general we can write,
\be 
W(\Delta E)=\frac{P(\Delta E)}{\tau(E)}\qquad ,
\label{gom15}
\ee
where $\tau(E)$ denotes the average time to escape from a configuration
of energy $E$. As these are microscopic transitions, the elementary escape process can be
assumed to be a Poisson process described by its characteristic time
$\tau(E)$. We can now imagine a situation where, in the large $N$ limit, the average escape
time is identical at both energies $E,E'$ whenever the energy difference
$\Delta E\sim {\cal O}(1)$. In this case, the timescale drops from the
ratio between the forward ($E\to E'$) and reverse ($E'\to E$)
transitions,
\be
\frac{W(\Delta E)}{W(-\Delta E)}=\frac{P(\Delta E)}{P(-\Delta E)}\qquad .
\label{gom16}
\ee
Therefore, the rates $W$ satisfy the same micorcanonical relation between the forward
and the reverse paths as do the transition probabilities $P(\Delta E)$.
Using \eq{gom14} and expanding the term in the exponential up to the
first order term in $\Delta E$ we find,
\bea
\frac{P(\Delta E)}{P(-\Delta E)}=\exp\Bigl
(S(E')-S(E)\Bigr)=\exp\Bigl[\Bigl(\frac{\partial S(E)}{\partial
E}\Bigr)\Delta E \Bigr]=\nonumber\\
\exp\Bigl[\frac{\Delta E}{T_{\rm eff}^{\rm FT}(t_w)} \Bigr]\qquad ,
\label{gom17}
\eea
where we have defined the time-dependent effective temperature,
\be
\frac{1}{T_{\rm eff}^{\rm FT}(t_w)}=\Bigl( \frac{\partial S(E)}{\partial E} \Bigr)_{E=E(t_w)}\qquad .
\label{gom18}
\ee
The effective temperature is a quantity that depends on the age of the
system $t_w$ through the time-dependent value of the energy $E(t_w)$.  A
remark is now in place. The decomposition \eq{gom15} is reminiscent of
the rates used in trap models~\cite{Bouchaud92}. However there is an
important difference between the GOM and trap models. In the latter the
timescale $\tau(E)$ also depends on the energy of the trap but,
contrarily to the present case, the rate is modified for energy changes
$\Delta E\sim {\cal O}(1)$. Moreover, the partial equilibration scenario
is difficult to visualize due to the absence of a proper
configurational space. Although phenomenological trap
models are very useful models to understand many aspects of the aging
dynamics, several issues still remain controversial specially regarding
the physical significance of FD violations~\cite{FieSol02,Ritort03,Sollich03}.

Relations describing ratios between transition rates for forward and
reverse processes in non-equilibrium systems are commonly known as
fluctuation theorems (FTs). The relations \eqq{gom16}{gom17} show a
strong resemblance with some of these theorems, however there are
important differences that we want to highlight. There are two general
classes of fluctuation theorems. In the first class there are the so
called entropy production FTs in stationary systems where the relation
between forward and reverse transitions is related to the entropy
production in the asymptotic limit of large times~\cite{EvaSea02}. In
the second class there are exact non-equilibrium work relations valid at all times
between the forward and reverse rates whenever the system is arbitrarily
perturbed away from an initial equilibrium state along both the forward
and reverse paths~\cite{Jarzynski97,Crooks98}.  The most important
difference between \eqq{gom16}{gom17} and these theorems concerns the
fact that the aging state is neither in an stationary state (first class)
and the system does not start from an initial equilibrium state (second
class). Moreover, the relations \eqq{gom16}{gom17} are not valid for energy
changes $\Delta E$ arbitrarily separated in time as the effective
temperature \eq{gom18} is age dependent.

There is another important feature of \eq{gom16} that must be
emphasized. Standard FTs allow for transitions to occur in both
directions and certainly this is what the identity \eq{gom16} seems to
imply.  Because
no work is exerted upon the system during the relaxation, the energy
change is related to the heat transferred between system and
bath. Therefore \eq{gom16} relates transition rates between identical
amounts of heat that are absorbed and released between the system and
bath. However, we face the problem that at zero temperature no heat can
be absorbed by the system (i.e the energy can never increase) and therefore \eq{gom16} cannot
hold. The resolution of this issue concerns the
true meaning of the effective temperature discussed in the next section. 

\subsection{Spontaneous relaxation and effective temperatures}
Developments during recent years in the theory of spin glasses and
glasses have shown that aging systems show violations of the
fluctuation-dissipation theorem~\cite{CriRit03} that can be quantified
in terms of an effective temperature. This is usually defined in terms
of the fluctuation-dissipation ratio (FDR)~\cite{CugKurPel97},
\be
T_{\rm eff}^{\rm FDR}(t,t_w)=\frac{1}{R(t,t_w)}\frac{\partial
C(t,t_w)}{\partial t_w},\qquad t>t_w
\label{gom19}
\ee
where we assume that $(t-t_w)/t_w\sim {\cal O}(1)$. In general the
effective temperature is a quantity that depends on the measured
observable and the probed frequency $\omega\sim 1/(t-t_w)$ relative to
the age, $\omega t_w=t_w/(t-t_w)$.  The interesting meaning of
\eq{gom19} is found in the low-frequency regime $\omega t_w<<1$ where
violations are expected to be strong. The effective temperature
defined in this way requires the measure of the response function,
i.e. the application of an external perturbation or field that shifts
the energy levels and exerts mechanical work upon the system. In this
case, the exerted work might account for part of the energy
transferred from the system to the bath making zero-field transitions
with $\Delta E>0$ accessible.  Therefore, \eqq{gom16}{gom17} have to
be understood as the proper way of quantifying forward and reverse
transitions in a configurational space that has been perturbed by the
action of an external field. This establishes a way to evaluate the
effective temperature from the rate of heat exchange between system
and bath {\em without the explicit need to introduce an applied
external field}. Note that an external field is usually required to evaluate the
response function $R(t,t_w)$~\footnote{In another context, numerical
methods to compute the response function have been recently
proposed~\cite{Chatelain03,Ricci03}.}.

The existence of the effective temperature is then related to the
presence of a heat exchange process between system and bath that we
call spontaneous as it is determined by the fact that the system has
been prepared in a non-equilibrium state.  The spontanoeus relaxation
is different from the heat exchange process (that we call stimulated)
between system and bath typical of equilibrium systems. In particular,
the stimulated process is a high-frequency process characterized a
Gaussian distribution of exchange events while the spontaneous process
is a low-frequency process that manifests in the form of some tails in
the heat-exhanged distribution which width is age
dependent~\footnote{The words ``spontaneous'' and ``stimulated'' make
explicit reference to the problem of light emission by atoms in a bath
of photons. In that case, the spontaneous process is the emission of
radiation by atoms in an excited state independently of the presence
of the bath. The stimulated process, though, is the emission and
absorption of energy by atoms induced by the bath of photons.}.  For a
recent discussion of these ideas see \cite{CriRit03b,Ritort03b}. The
existence of these two heat exchange processes is at the root of the
intermittency phenomenon recently observed in glasses and
colloids~\cite{Cil03,Cip03}.

The effective temperature can be computed in the GOM by using \eq{gom17}
and \eq{gom18}. Indeed, from \eq{gom17} and \eq{gom5} we get,
\be
\frac{P(\Delta E)}{P(-\Delta E)}=\exp\Bigl(\frac{2M_{\Delta E}\Delta
E}{\sigma_{\Delta E}^2}\Bigr)\qquad ,
\label{gom20}
\ee
and therefore,
\be
T_{\rm eff}^{\rm
FT}(t_w)=\lambda(t_w)=\frac{\overline{(V^{'}(x))^2}}{\overline{V^{''}(x)}}\qquad ,
\label{gom21}
\ee
where $\lambda$ is an age dependent quantity that has been defined in
\eq{gom9} and where we have substituted \eqq{gom6a}{gom6b}. In equilibrium it is
straightforward to check that \eq{gom21} coincides with the bath
temperature by integrating by parts twice the integral in the
denominator of \eq{gom21},
\be
\overline{V^{''}(x)}=\int_{-\infty}^{\infty}dx V^{''}(x) \exp(-\beta
V(x))=\beta\int_{-\infty}^{\infty}dx (V^{'}(x))^2 \exp(-\beta
V(x))\qquad .
\label{gom22}
\ee
In the aging regime in a partial equilibration scenario the effective
temperature \eq{gom21} is related to the energy at time $t_w$ through
the relation \eq{gom18}.

\section{The homogeneous potential model}
\label{homogeneous}
An example of the GOM where many quantities can be easily
worked out is the case of an homogeneous potential of the type,
\be
V_p(x)=\frac{k}{2p}x^{2p}\qquad ,
\label{gom23}
\ee
with $p$ a positive integer. The probability distribution \eq{gom5} is
given by,
\be
P(\Delta E)=(2\pi
k^2\Delta^2h_{2p})^{-\frac{1}{2}}\exp\Bigl(-\frac{(\Delta
E-\frac{k(2p-1)}{2}\Delta^2 h_p)^2}{2k^2\Delta^2 h_{2p}} \Bigr)\qquad ,
\label{gom24}
\ee
with $h_k=\overline{x^{2(k-1)}}$. By definition
$E=N\overline{V_p}=N(k/2p)h_{p+1}$. Similar results to \eq{gom24} can
then be obtained for the distribution of changes $P(\Delta h_k)$ for
generic observables $h_k$ which lead to a hierarchy of coupled
dynamical equations similar to \eq{gom8a}. These equations can then be
studied using generating functional techniques similar to those
developed in other solvable spin-glass models~\cite{Ritort96}. Only
for the harmonic case $p=1$ the equation for the energy \eq{gom8a} is
closed and Markovian as its time evolution only depends on the energy.

The expression for the effective temperature \eq{gom21} for the
homogeneous potential model is given by,
\be
T_{\rm eff}^{\rm FT}(t_w)=\frac{k h_{2p}(t_w)}{(2p-1)h_p(t_w)}\qquad .
\label{gom25}
\ee
In the partial equilibration scenario all observables are functions of
the energy $E$ of the hypersurface over which the system partially equilibrates.
The relation between the value of $T_{\rm eff}^{\rm FT}(t_w)$ and the
energy $E(t_w)$ can be easily derived from \eq{gom18}. Introducing
\eq{gom23} in \eq{gom2} we get,
\be
{\cal Z}_1=\int_{-\infty}^{\infty} \exp\Bigl( -\beta
V_p(x)\Bigr)=\Bigl(\frac{2p}{\beta k}
\Bigr)^{\frac{1}{2p}}\int_{-\infty}^{\infty}dy \exp(-y^{2p})\qquad ,
\label{gom26}
\ee
yielding the free energy $F={\rm const}-NT\log({\cal Z}_1)={\rm
const}-(NT/2p)\log(T)$ and the following expressions for the energy and
entropy,
\bea
E=\frac{\partial \beta F}{\partial \beta}=\frac{NT}{2p}\qquad ,\label{gom27a}\\
S=-\frac{\partial F}{\partial
T}=\frac{N}{2p}+\frac{N\log(T)}{2p}=\frac{N\log(E)}{2p}+{\rm const}\qquad ,\label{gom27b}
\eea
giving,
\be
\frac{1}{T_{\rm eff}^{\rm FT}(t_w)}=\Bigl( \frac{\partial S(E)}{\partial E} \Bigr)_{E=E(t_w)}=\frac{N}{2pE(t_w)}\qquad .
\label{gom28}
\ee
The case $p=1$ corresponds to the harmonic oscillator and gives the well
known equipartition relation $T_{\rm eff}^{\rm FT}(t_w)=2E(t_w)$. As
remarked in the paragraph following \eq{gom9c} the dynamical evolution of the energy can be
solved in general by knowing the relation \eq{gom28} between the energy
$e=E/N$ and the effective temperature $T_{\rm eff}^{\rm FT}(t_w)=\lambda(t_w)$. In this case it is
possible to derive the following asymptotic behavior for the energy,
\be
e(t)\sim \frac{1}{(\log(t))^p}~~~+~~~ {\rm subleading\,\,\, logarithmic\,\,\, corrections}\qquad . 
\label{gom28b}
\ee
The effective temperatures \eq{gom25} can be also derived using the
fluctuation-dissipation relation \eq{gom19} or a set of microcanonical
relations describing observable changes. The possibility to obtain the
same value of the effective temperature by using three different
approaches (the FDR \eq{gom19}, the FT \eq{gom20} and the microcanonical
rates for observables) has been explicitly shown for the harmonic case
$p=1$~\cite{Ritort03b}.  In Ref.~\cite{Ritort03b} the main assumption
was the validity of the partial equilibration scenario. Because the
partial equilibration scenario also holds for the GOM, the main
conclusions of Ref.~\cite{Ritort03b} are expected to hold also for the
present more general case.  

The procedure to derive the FDR
\eq{gom19} entails the computation of the correlation and response
functions for the magnetization $M$ defined as $M(t)=\sum_i x_i(t)$,
\bea
C(t,t_w)=\frac{1}{N}\sum_{i=1}^N \overline{x_i(t)x_i(t_w)}\qquad ,\label{gom29a}\\
R(t,t_w)=\overline{\frac{\delta x(t)}{\delta h(t_w)}}\qquad , \label{gom29b}
\eea
where $h(t_w)$ is an impulse field coupled to the magnetization $M$ at
time $t_w$. 
At $T=0$ the response \eq{gom29b} is finite due to the shift of the
energy levels induced by the field. For the harmonic case
$p=1$, dynamics is closed and a simple expression can be derived for the
effective temperature \eq{gom19},
\be
T_{\rm eff}^{\rm FDR}(t,t_w)=2e(t_w)+\frac{1}{f(t_w)}\frac{\partial
e(t_w)}{\partial t_w}\qquad ,
\label{gom30}
\ee
where $f(t_w)$ is a function that asymptotically decays as $1/t_w$. Two
remarkable facts emerge from \eq{gom30}: 1) Eq.\eq{gom30} only depends
on the age $t_w$ at all times $t>t_w$, therefore $T_{\rm eff}^{\rm
FDR}(t,t_w)\equiv T_{\rm eff}^{\rm FDR}(t_w)$ characterizes the aging
state of the system at time $t_w$; 2) The second term in the r.h.s of
\eq{gom30} is subdominant respect to the first term. Using \eq{gom28}
this leads to $T_{\rm eff}^{\rm FDR}(t_w)\to T_{\rm eff}^{\rm
FT}(t_w)\to 2e(s)$ so both the effective temperature derived from the
FDR and the FT coincide.

\begin{figure}
  \centering \includegraphics[scale=0.3]{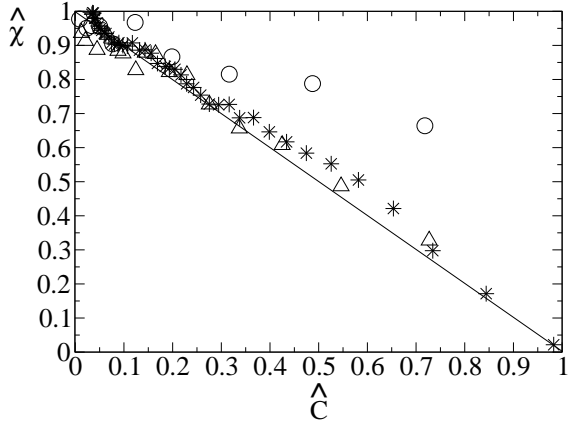} 
  \caption{FD plot for the harmonic model ($p=1$) obtained from Monte
  Carlo simulations at $T=0$ with $N=1000$ oscillators, $k=\Delta=1$,
  magnetic field intensity equal to 0.01 and $t_w=1,10,100$
  (circles,triangles up and stars respectively). It represents
  $\hat{\chi}(\hat{C})$ and the straight line is the theoretical
  asymptotic prediction for large values of $t_w$. Data were averaged
  over 200 dynamical histories.}
\label{fig3}
\end{figure}

The obtain the effective temperature it is sometimes useful to
construct the so-called fluctuation-dissipation (FD)
plots~\cite{CugKur94,FraRie95,SolFieMay02,CriRit03}. The FD plots for
the homogeneous model can be worked out as follows (the same
construction holds for the GOM). As
$C(t_w,t_w)=\overline{x^2(t_w)}=h_2(t_w)$ is time dependent then it is
convenient to normalize the correlation $C(t,t_w)$ by the
autocorrelation value taken at the lowest time
$\hat{C}(t,t_w)=C(t,t_w)/C(t_w,t_w)$ and plotting the
integrated-response $\chi(t,t_w)=\int_{t_w}^tdsR(t,s)$ as a function
of $\hat{C}$ for $t_w$ fixed and varying $t$. The resulting asymptotic curve is
then expected to have a the form of a straight line,
$\hat{\chi}(\hat{C})=\hat{\chi}_0(1-\hat{C})$ where $\chi_0$ is the
equilibrium susceptibility at zero temperature $\hat{\chi}_0\propto
\beta^{\frac{(p-1)}{p}}$. These straight FD plots are characteristic
of the one-step behavior observed in structural
glasses. Fig.~\ref{fig3} shows the resulting FD plot for the $p=1$
case as obtained from numerical simulations of the model.

\section{The wedge potential model}
\label{wedge}
\begin{figure}
  \centering \includegraphics[scale=0.3]{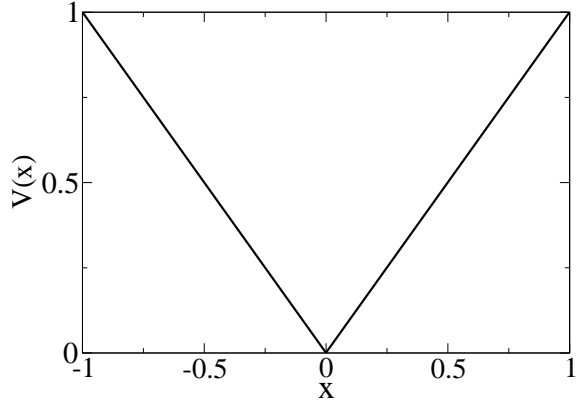} 
  \caption{The wedge potential model \eq{gom31} with
  $k=1$. It can be seen as a special case of the homogeneous model of
  Sec.~\protect\ref{homogeneous} with $p=1/2$.}  
  \label{fig4}
\end{figure}

An interesting example of the GOM is the case where the first
derivative of the potential $V'(x)$ is not continuous at the ground
state configuration $x=0$. In this case the expressions \eqq{gom6a}{gom6b} need
to be reconsidered. The classical example for a potential of this type is the 
wedge potential model defined by,
\be
V(x)=k|x|\qquad .
\label{gom31}
\ee
This potential is depicted in Fig.~\ref{fig4}. The statics for this
model is straightforward and the partition function is given by ${\cal
Z}_1=2/(\beta k)$. The internal energy and entropy are given by the
expressions \eqq{gom27a}{gom27b} with $p=1/2$. At equilibrium the
one-oscillator probability density $q^{\rm eq}(x)$ is given by,
\be
q^{\rm eq}(x)=\frac{k\beta}{2}\exp(-\beta k |x|)\qquad .
\label{gom32}
\ee

\begin{figure}
  \centering \includegraphics[scale=0.5]{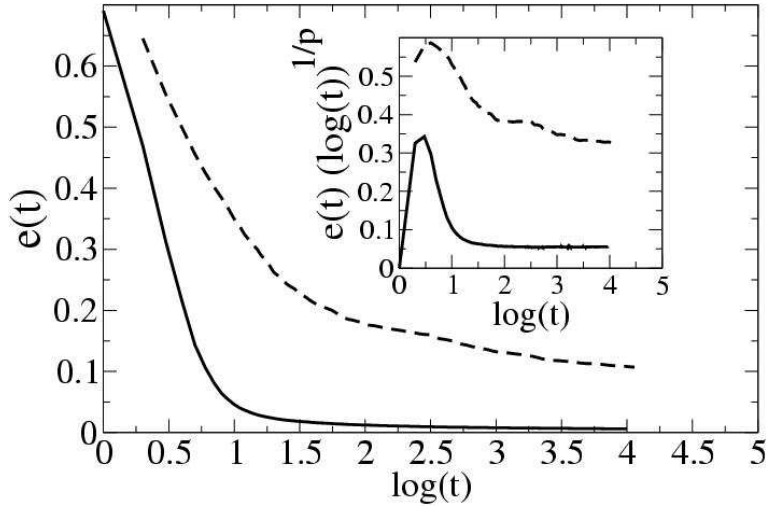} \caption{Monte Carlo
  studies of the harmonic model (continuous line) and the wedge model
  (dashed line). Simulations have been done at $T=0$ with $N=1000$ oscillators and
  $k=\Delta=1$. Main: Energy decay $e(t)$ in both models. Inset: e(t)
  multiplied by $(\log(t))^{1/p}$ (harmonic model with $p=1$, wedge
  model with $p=1/2$) as a function of $\log(t)$. In the large $t$ limit
  both curves converge to a constant value.}  \label{fig5}
\end{figure}

\begin{figure}
\vspace{1cm}
  \centering \includegraphics[scale=0.4]{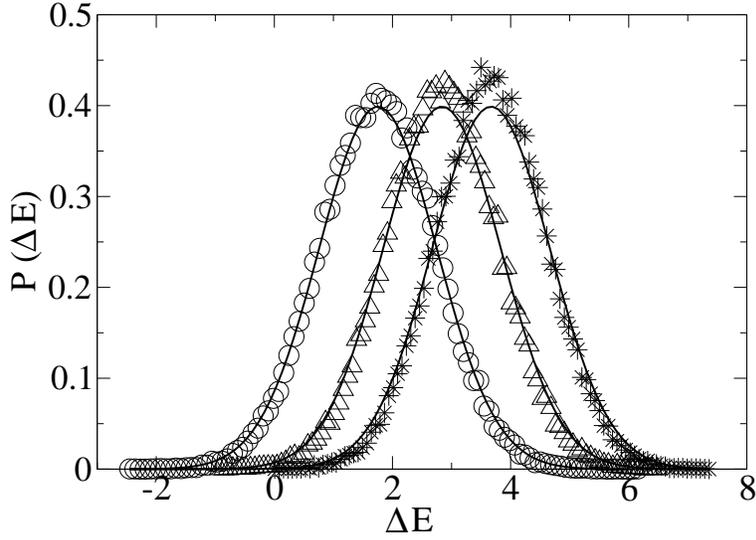} 
  \caption{Probability distribution $P(\Delta E)$ \eq{gom37} for the wedge
  model numerically evaluated at $T=0,N=1000,k=\Delta=1$ for
  $t_w=10,100,1000$ (circles,triangles up and stars respectively). The
  continuous lines are the fitted Gaussians with $M_{\Delta
  E}=q_{t_w}(0)$ as fitting parameter. The effective temperature
  \eq{gom38} is then compared with that obtained from the energy in
  Table I.}  
  \label{fig6}
\end{figure}

\begin{figure}
 \vspace{.5cm}
 \centering \includegraphics[scale=0.4]{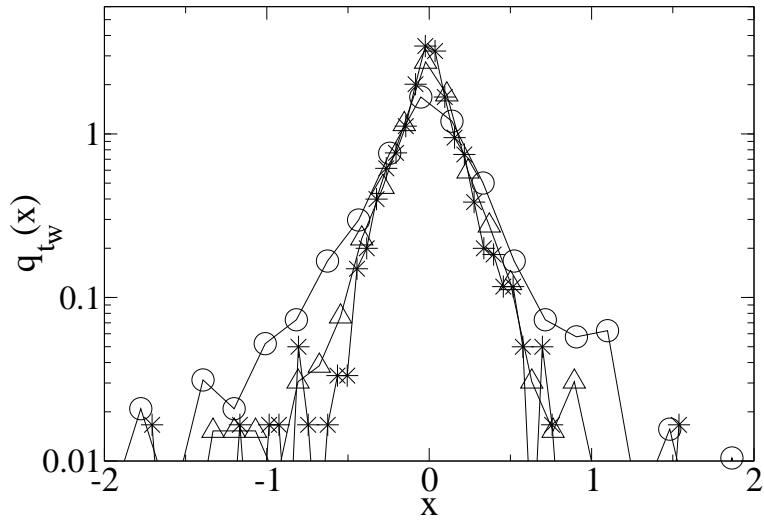}
  \caption{One-oscillator probability distribution $q_{t_w}(x)$. The
  same parameters and symbols as in Fig.~\protect\ref{fig6}. Note the
  presence of the exponential tails in the distribution as expected
  from \eq{gom38b}.}
  \label{fig7}
\end{figure}

To solve the off-equilibrium dynamics of this model we proceed similarly
as was done in Sec.~\ref{sec:gom} for a general function $V(x)$. The
main difference now is that the potential \eq{gom31} is not
differentiable at $x=0$. While the expression for $\sigma_{\Delta E}^2$
is well defined ($\overline{(V'(x))^2}=k^2$ is continuous for all $x$)
the expression for $M_{\Delta E}$  is not as the second derivative
$\overline{V''(x)}$ is discontinuous at $x=0$. The expression however
can be guessed by noticing that $V''(x)=a\delta(x)$ and using the
relation,
\be
\int_{-\infty}^{\infty} dx V''(x)=a=V'(\infty)-V'(-\infty)=2k\qquad .
\label{gom33}
\ee
From \eqq{gom6a}{gom6b} this gives,
\be
M_{\Delta E}=k\Delta^2 \overline{\delta(x)}=k\Delta^2 q_t(0)\qquad ,
\label{gom34}
\ee
where $q_t(x)$ is the one-oscillator probability density,
\be
q_t(x)=\frac{1}{N}\sum_{i=1}^N\delta(x-x_i(t))\qquad .
\label{gom35}
\ee
The result \eq{gom34} together with the relation,
\be
\sigma_{\Delta E}^2=k^2\Delta^2\qquad ,
\label{gom36}
\ee
gives the final probability distribution,
\be
 P(\Delta E)=\bigl(2\pi k^2\Delta^2\bigr)^{-\frac{1}{2}}
      \exp\left[-\frac{(\Delta E-k\Delta^2 q_t(0))^2}{2
 k^2\Delta^2}\right] \qquad .
\label{gom37}
\ee
This result can be alternatively derived doing a more elaborated but
better controlled calculation that we do not consider
interesting enough to reproduce here in detail.
The effective temperature \eq{gom19} can be obtained using the FT
\eqq{gom20}{gom21},
\be
T_{\rm eff}^{\rm FT}(t_w)=\frac{k}{2q_{t_w}(0)}\qquad .
\label{gom38}
\ee
Using the thermodynamic relation \eq{gom27a} with $p=1/2$ we get
$e=E/N=k\overline{|x|}=T$. If we now assume that \eq{gom32}
holds in the aging regime by replacing $\beta$ with $\beta_{\rm
eff}(t_w)=1/T_{\rm eff}(t_w)$ then, in a partial equilibration scenario
we get, 
\be
q_{t_w}(x)=\frac{k\beta_{\rm eff}(t_w)}{2}\exp(-\beta_{\rm eff}(t_w) k |x|)\qquad ,
\label{gom38b}
\ee
which gives $q_{t_w}(0)=(k\beta_{\rm eff}(t_w))/2=k/(2e(t_w))$. Substituting
this last result in \eq{gom38} we obtain,
\be
T_{\rm eff}^{\rm FT}(t_w)=\frac{k}{2q_{t_w}(0)}=e(t_w)\qquad ,
\label{gom39}
\ee
which coincides with the result derived using the thermodynamic
relations \eq{gom18} and \eq{gom27b} with $p=1/2$. Finally, we mention
that the asymptotic decay for the energy derived for the homogeneous
potential model \eq{gom28b} also holds in the wedge model substituting $p=1/2$,
i.e. $e(t)\sim 1/\sqrt{\log(t)}$. In
Figs.~\ref{fig5},\ref{fig6},\ref{fig7} we show some numerical results
for the wedge model obtained by doing Monte Carlo calculations. These
have been done at zero temperature for $N=1000$ oscillators with
parameters $k=\Delta=1$ and starting from a random initial configuration
with initial coordinates $x_i$ chosen from a Gaussian distribution of
zero mean and unit variance. These simulations are useful to verify the
main predictions. Fig.~\ref{fig5} shows the time decay of the energy for
the wedge model compared to the harmonic model. Fig.~\ref{fig6} is a
test of the main result \eq{gom37}. The energy change distributions have
been evaluated at three different waiting times $t_w=10,100,1000$ and
fitted to a Gaussian distribution of variance $k^2\Delta^2=1$ where the
mean $q_{t_w}(0)$ is the fitting parameter. This value is then compared
with the value of the energy $e(t_w)$ to check the identity
\eq{gom39}. The results are shown in table I.  Fig.~\ref{fig7}
shows the one-oscillator probability density in the aging regime
$q_{t_w}(x)$ at three different waiting times compared to the expected
result \eq{gom38b}.

\begin{table}[b]
\begin{center}
\begin{tabular}{|c|c|c|}\hline
$t_w$ & $T_{\rm eff}^{\rm FT}(t_w)=e(t_w)$ & $T_{\rm eff}^{\rm FT}(t_w)=k/(2 q_{t_w}(0))$ \\\hline
 10 & 0.335 & 0.28 \\\hline
 100 & 0.180 & 0.178 \\\hline
 1000 & 0.132 & 0.136 \\\hline
 \end{tabular}
\caption{Effective temperature in the wedge model \eq{gom39} obtained in
two independent ways. Second column: From the value of the energy
$e(t_w)$ shown in  Fig.~\ref{fig5}. Third column: 
From the value of $q_{t_w}(0)$ obtained by fitting the expression
\eq{gom37} to the numerical distribution $P(\Delta E)$ shown in
Fig.~\protect\ref{fig6}. Both values asymptotically
coincide for large values of $t_w$.}
\label{table}
\end{center}
\end{table}


\section{Conclusions}
In this paper we have introduced a new family of exactly solvable models
characterized by zero-temperature relaxation determined by entropy
barriers and partial equilibration.  Generalized oscillator models
(GOMs) offer a conceptual framework to develop a statistical description
of the aging state. The interesting property of this class of models
stems from the validity of the partial equilibration
scenario~\cite{Ritort03b}: dynamics is ergodic and microscopically
reversible when the system is constrained to move along the constant
energy hypersurface.  We have then computed the probability distribution
of energy changes $P(\Delta E)$ that characterizes the spontaneous
relaxation process at zero temperature. The spontaneous process is not
thermally activated but determined by the fact that the system has been
prepared in a non-equilibrium aging state.  Using a fluctuation theorem
for the aging state it is then possible to derive analytic expressions
for the effective temperature \eq{gom21} without the need to solve the
dynamical equations for correlations and responses \eq{gom19}. The
quantitative description of the spontaneous process in terms of a
fluctuation theorem valid in the aging state has been recently
proposed~\cite{CriRit03b} to be at the root of the intermittency
phenomenon observed in glasses and colloids~\cite{Cil03,Cip03}.

Two classes of models have been studied in detail. The homogeneous
potential model in Sec.~\ref{homogeneous} and the wedge potential model
in Sec.~\ref{wedge}. Particularly interesting is the latter where
the first derivative of the potential is singular at the ground state configuration. In this
case, the effective temperature \eq{gom38} depends on the value of the
one-oscillator probability distribution at the value of the singularity
of the potential. The present studies can be extended to other
interesting potential functions and some results will be presented in the future. 

A salient feature of the GOM is the Gaussian shape of the distribution
$P(\Delta E)$ in large $N$ limit. This is a direct consequence of the
dynamics of the model \eq{gom3} which is of the mean-field type as
oscillator correlations do not enter the analytical expression of $P(\Delta E)$. In this
regard, the validity of the partial equilibration scenario in the
GOM is consequence of the mean-field character of the dynamics. The
extension of these ideas to non mean-field dynamical models is an open
problem. Nevertheless, despite of the fact that the present ideas have
been derived from the study of mean-field systems, we do not foresee conceptual limitations
in their adaptation to spatially correlated dynamics. A
conceptual description of the aging state in terms of heat exchange
processes \cite{CriRit03b,Ritort03b} could be achieved in terms of a
spatially fluctuating effective temperature that would describe local
fluctuations in the rate of energy relaxation in the system. At
difference with bath temperatures, effective temperatures have to be
considered as fluctuating intensive variables as they describe energy
exchange processes (either heat releasing or work releasing -through
mechanical stresses-) occurring over nanoscale spatial regions where
energies are not macroscopic but of the order of few $k_BT$. Would be
very interesting to explore the possible connection between these ideas and the
existence of spatial heterogeneities that have received considerable
attention during the past years (see for instance
\cite{GarCha03,BerGar03}). Establishing a thermodynamic
description of these heterogeneous excitations is probably an important step
towards their understanding.

{\bf Acknwoledgements.} The author acknowledges financial support from
the David and Lucile Packard foundation, the Spanish Ministerio de
Ciencia y Tecnolog\'{\i}a Grant BFM2001-3525, the Catalan government,
the STIPCO EEC network and the SPHINX ESF program.

\end{document}